\documentclass[twocolumn,showpacs,amssymb]{revtex4}
\usepackage{graphicx}
\usepackage{dcolumn}
\usepackage{bm}

\def\beq{\begin{equation}}
\def\enq{\end{equation}}
\def\ba{\begin{eqnarray}}
\def\ea{\end{eqnarray}}
\def\<{<\!\!}
\def\>{\!\!>}

\begin{document}
\input{epsf}

\title{Testing the Equivalence Principle and Lorentz Invariance with PeV Neutrinos from Blazar Flares }

\author{Zi-Yi Wang$^{1,2}$, Ruo-Yu Liu$^{3}$ and Xiang-Yu Wang$^{1,2*}$ }

\affiliation{$^1$School of Astronomy and Space Science, Nanjing University, Nanjing 210093, China\\
$^2$ Key laboratory of Modern Astronomy and Astrophysics (Nanjing
University), Ministry of Education, Nanjing 210093, China\\
$^{3}$Max-Planck-Institut f\"ur Kernphysik, 69117 Heidelberg,
Germany\\
*Electronic address: xywang@nju.edu.cn }

\begin{abstract}

It was recently proposed that a giant flare of the blazar PKS
B1424-418 at redshift $z=1.522$ is in  association with  a
PeV-energy neutrino event detected by IceCube. Based on this
association  we here suggest that the flight time difference
between the PeV neutrino and gamma-ray photons from blazar flares
can be used to constrain the violations of equivalence principle
(EP) and the Lorentz invariance for neutrinos. From the calculated
Shapiro delay due to clusters or superclusters in the nearby
universe, we find that violation of the equivalence principle for
neutrinos and photons is constrained to an accuracy of at least
$10^{-5}$, which is two orders of magnitude tighter than the
constraint placed by MeV neutrinos from supernova 1987A. Lorentz
invariance violation (LIV) arises in various quantum-gravity
theories, which predicts an energy-dependent velocity of
propagation in vacuum for photons and neutrinos. We find that the
association of the PeV neutrino with the gamma-ray outburst set
limits on the energy scale of possible LIV to $> 0.01 E_{pl}$ for
linear LIV models and $>6\times 10^{-8} E_{pl}$ for quadratic
order LIV models, where $E_{pl}$ is the Planck energy scale. These
are the most stringent constraints on neutrino LIV for subluminal
neutrinos.

\end{abstract}

 \maketitle

{\em Introduction---} The IceCube Collaboration recently announced
the discovery of extraterrestrial neutrinos, including a couple of
PeV-energy scale neutrinos \cite{IceCube2013, IceCube2014}. The
third PeV neutrino (IC 35) has an energy of
$2004^{+236}_{-262}{\rm TeV}$ and a median positional uncertainty
of $R_{50} = 15.9^\circ$ centered at the coordinate
RA=$208.4^\circ$, Dec=$-55.8^\circ$ (J2000). By searching this
field for positional coincidences with gamma-ray-emitting active
galactic nuclei (AGNs), a high-fluence outburst of the blazar PKS
B1424-418 is found to be in temporal and positional coincidence
with the neutrino event\cite{Kadler2016}.   An association between
the giant flare of PKS B1424-418 and the PeV neutrino is then
suggested based on the unprecedented nature of the two events and
a posteriori probability for chance coincidence of about $5\%$
(i.e. a 2 $\sigma$ confidence correlation)\cite{Kadler2016}. It is
also argued that the flare provides an energy output high enough
to explain the observed PeV event\cite{Kadler2016}. PKS B1424-418
is at redshift z = 1.522 and classified as a flat spectrum radio
quasar\cite{White1988}.

Assuming a real   association between the outburst activity of PKS
B1424-418 and the PeV neutrino, we now use the flight time
difference between the PeV neutrino and the blazar photons to
constrain violations of the EP and  Lorentz invariance, since both
theories predict  flight time difference for particles of
different-species or with different energies. PKS B1424-418 showed
a long-lasting high-fluence outburst in 100 MeV to 300 GeV
gamma-rays over the period from 2012 Jul 16 to 2013 Apr 30, which
spans the arrival time of the PeV neutrino. The maximum possible
time-travel delay between the beginning of the outburst and the
arrival of the neutrino is about 160 days\cite{Kadler2016}.

The Einstein equivalence principle (EP) states that trajectory of
any freely  falling uncharged test body is independent of its
internal composition or structure. It is a fundamental postulate
of general relativity as well as other metric theories of
gravity\cite{Will}. Delays between arrival times of photons and
neutrinos from SN 1987A has been used to test EP for neutrinos and
photons through the effect of Shapiro (gravitational) time
delay\cite{Longo1988,Krauss&Tremaine1988}. The near simultaneous
arrival of photons and neutrinos from this core-collapse supernova
confirmed that the Shapiro delay for neutrinos is the same as that
for photons to within 0.2-0.5\%. Recently, such a test has been
applied to different-energy photons in extragalactic transients
(e.g. gamma-ray bursts, fast radio bursts, TeV
blazars)\cite{Gao2015,Wei2015,Wei2016,Zhang2016,Nusser2016}, and
also to gravitational waves\cite{Wu2016,Kahya2016}.

Some quantum-gravity (QG) models postulate  an  inherent structure
of spacetime (e.g.  a foamy structure) near the Planck energy
scale ($E_{pl}=\sqrt{(\hbar c^5)/G}\simeq1.22\times10^{19}{\rm
GeV}$) (see e.g. ref.\cite{Amelino-Camelia1998} for a review). It
is speculated that such QG effects can possibly lead to violations
of Lorentz Invariance (LIV) at very high
energies\cite{Amelino-Camelia1998}. One manifestation of such
violations could be a velocity dispersion, in which the speed of a
photon or neutrino in vacuum becomes dependent on its energy.
Currently, the most stringent limits (at 95\% CL) on the "QG
energy scale" (the energy scale that LIV-inducing QG effects
become important, $E_{QG}$) are obtained from the GeV photons in
the gamma-ray burst (GRB) 090510. The limits are $E_{QG,1}
> (1-10) E_{pl}$ and $E_{QG,2}
> 10^{-8}E_{pl}$ for linear and quadratic leading order
LIV-induced vacuum dispersion,
respectively\cite{Abdo2009,Vasileiou2013}.    Using the SN1987a
neutrino data, ref. \cite{Ellis2008} set limits of $E_{\nu,QG1}
> 2.7 \times10^{-9}E_{pl}$ and $E_{\nu,QG2} > 4
\times10^{-15}E_{pl}$ for linear and quadratic order LIV models
respectively. Jacob \& Piran\cite{Jacob2007} proposed that GRB
neutrinos at cosmological distance, if detected, can constrain LIV
to a much higher precision. Although so far no neutrinos have been
detected from GRBs\cite{Aartsen2015}, future detection would make
this approach very useful.

{\em Constraints on  EP--} The motion of neutrinos and photons in
a gravitational field can be described with the post-Newtonian
(PPN) formalism. Each gravity theory satisfying EP is specified by
a set of PPN parameters. The limits on the differences in PPN
parameters for different particles describe the accuracy of EP.
The time interval that photons or neutrinos require to traverse a
given distance is longer by
\begin{equation}
\delta t=-\frac{1+\gamma}{c^3}\int_{e}^{a}U(r)\rm{d}r,
\end{equation}
in the presence of a gravitational potential
${U}(r)$\cite{Shapiro1964}, where $e$ and $a$ denote the location
of the emission and arrival of the particles. $\gamma$ is the
parameterized PPN parameter, which is $\gamma=1$ in general
relativity.

To calculate the time delay with Eq.(1), we have to figure out the
gravitational potential.  The total gravitational potential
consists of three parts, respectively contributed  by the blazar's
host galaxy, the intergalactic space and our Milky Way. For
sources at cosmological distances, the Shapiro delay due to nearby
clusters and/or superclusters turns out to be more important than
the  Milky Way and AGN host galaxy\cite{Minakata1996PhRvD}, as
will be shown below.

One of the nearest clusters to the Milky Way is the Virgo Cluster.
The center of Virgo  Cluster lies at a distance $\approx16.5$
Mpc\cite{Mei2007}  at the position of RA=$186.8^\circ$ and
Dec=$12.7^\circ$ (J2000). Its mass is estimated to be
$1.2\times10^{15}\ \rm{M_\odot}$ within $8$ degrees ($2.2$ Mpc)
from the center of the cluster\cite{Fouque2001}. The angle between
the  direction of PKS B1424-418 and the Virgo Cluster is
$\approx60^\circ$, much greater than the radius of the Virgo
Cluster, so the cluster can be treated as a point source when
calculating the gravitational potential.

The time delay is given by\cite{Longo1988}
\begin{equation}
\delta
t=(1+\gamma)M_c\rm{ln}\{[X_S+(X_S^2+b^2)^{1/2}][X_D+(X_D^2+b^2)^{1/2}]/b^2\},
\end{equation}
where $M_c$ is the mass of the gravitational field source (the
Virgo Cluster), $X_S$  is the projected distance between the
particle source (PKS B1424-418) and the gravitational source on
the line of sight of PKS B1424-418, $X_D$ is the projected
distance of the gravitational field source on the line of sight of
PKS B1424-418, and $b$ is the impact parameter (see Fig.1 in ref.
\cite{Longo1988}). Since in our case $X_S$ is approximately equal
to the distance of the blazar at $z=1.522$, we take $X_S\approx
4400\ \rm{Mpc}$, which is much larger than $b$. The deflection
angle of light is much less than the angle between the directions
of the particle source and the gravitational field source, $\delta
\theta \ll \theta$, thus Eq. (2) can be simplified as
\begin{equation}
\delta
t=(1+\gamma)M_c\rm{ln}(\frac{2}{1-\cos\theta}\frac{X_S}{d}),
\end{equation}
where $d$ is the distance of the gravitational field source.
Noting that the logarithmic dependence on the angle and  the
distance of the gravitational source,  the time delay is mostly
determined by the mass of the gravitational field source.
Therefore, the Shapiro delay due to nearby clusters  in the
intergalactic space is more important than the  Milky Way and AGN
host galaxy. For the clusters and superclusters around the
direction of the blazar,
$\rm{ln}(\frac{2}{1-\cos\theta}\frac{X_S}{d})\sim 5$. For
$\gamma\simeq1$ the time delay caused by the Virgo Cluster is
about $8.2\times10^{10}$ s.

The Great Attractor is another  mass concentration in the nearby
universe. It has  a mass of $\sim 5\times 10^{16}\
\rm{M_\odot}$\cite{Lynden-Bell1988}  and locates at a distance of
about $63$ Mpc. Its position is at about RA=$200^\circ$ and
Dec=$-54^\circ$ (J2000), so $\theta\approx17^\circ$. The Shapiro
time delay due to the Great Attractor is then $4.0\times10^{12}$
s.

Since the maximum possible time-travel delay between the beginning
of  the outburst and the arrival of the neutrino is about 160
days, from the relation
\begin{equation}
\frac{\gamma_\nu-\gamma_\gamma}{\gamma+1}= \frac{\delta
t_\nu-\delta t_\gamma}{\delta t},
\end{equation}
we obtain $\gamma_\nu-\gamma_\gamma \leq 3.4\times10^{-4}$  and
$\gamma_\nu-\gamma_\gamma \leq 7.0\times10^{-6}$, respectively,
for the Virgo Cluster and Great Attractor. The limit obtained with
the Great Attractor is  about 2-3 orders of magnitude tighter than
the previous limits obtained with MeV neutrinos from SN1987A.

{\em Constraints on LIV.--} The LIV predicts an energy-dependent
velocity of propagation in vacuum for particles. Taking into
account only the leading order correction, we expect, for
particles with $E\ll E_{QG}$, an approximate dispersion relation
\begin{equation}
E^2\simeq p^2c^2+m^2c^4\pm E^2(\frac{E}{E_{QG}})^n,
\end{equation}
where $c$ is the constant speed of light (at the limit of zero
photon energy), and +1 (-1) corresponds to the "subluminal" (
"superluminal") case. Noting that the superluminal neutrinos will
produce electron-positron pairs and loss their energy
rapidly\cite{Cohen2011}, very stringent constraints on LIV have
been obtained for superluminal
neutrinos\cite{Borriello2013,Stecker2014,Diaz2014}. We here
consider the constraints for subluminal neutrino LIV. For
$E<E_{QG}$ , the lowest order term in the series is expected to
dominate the sum. In the case that the n = 1 term is suppressed,
something that can happen if a symmetry law is involved, the next
term n = 2 will dominate\cite{Vasileiou2013}.  Here we only
consider the n = 1 and n = 2 cases, since the data are not
sensitive to higher order terms for $E<E_{QG}$. The limit on
$E_{QG,n}$ for $n>2$   would be less constraining than the case
with a lower  n since the delay time is proportional to
$(\frac{E}{E_{QG,n}})^n$. The  LIV time delay  of a high energy
neutrino with an observed energy, $E$, emitted at redshift  $z$
is\cite{Jacob2007}
\begin{equation}
\Delta
t=\frac{1+n}{2H_0}\frac{E^n}{E_{QG}^n}\int_0^z\frac{(1+z')^n}{\sqrt{\Omega_\Lambda+\Omega_M(1+z')^3}}dz',
\end{equation}
where $H_0=67.8 \rm km/s/Mpc$ is the Hubble constant and
$\Omega_\Lambda=0.692$, $\Omega_M=0.308$ obtained by the Planck
collaboration\cite{Planck2015}. The maximum possible time-travel
delay between the beginning of the gamma-ray outburst and the
arrival of the neutrino is $\sim160$ days, so we take $\Delta
t=160 {\rm d}$ as an upper limit of the delay between photons and
the PeV neutrino. For linear LIV (n = 1), we obtain a limit on the
"QG energy scale" as $E_{QG,1}>0.009E_{pl}$. For the quadratic
order LIV-induced vacuum dispersion, we obtain
$E_{QG,2}>6\times10^{-8}E_{pl}$. Although the constraints on the
linear LIV models is not as stringent as that placed by GeV
photons from GRB~090510, the constraints on the quadratic  order
LIV models is the most stringent so far. As a comparison, the
previously published most stringent limits for the quadratic order
LIV is obtained from the bright  flares of PKS 2155-304 observed
by HESS ($E_{QG,2}>5\times10^{-9}E_{pl}$) and the Fermi GRB 090510
($E_{QG,2}>10^{-8}E_{pl}$)\cite{Abramowski2011,Vasileiou2013}.
Thus, the constraints placed by the blazar PeV neutrino on the
quadratic order LIV is a factor of 6 tighter than previous
constraints. Moreover, our constraints for both linear and
quadratic  order   models are the most stringent limits on
neutrino LIV for  subluminal neutrinos.

{\em Discussions--} Assuming a physical association  between the
PeV neutrino and the giant outburst of blazar  PKS B1424-418, we
constrained the violation of the EP and LIV using the arrival time
delay between the neutrino and photons. We note that a 5 \%
probability for a chance coincidence between the giant outburst
and PeV neutrino remains and our results are based on  the
suggestion that the association is physical. Blazar flares have
long been suspected to be able to produce high-energy neutrinos
(e.g. refs.\cite{Stecker1991,Mannheim1992}; see
ref.\cite{Murase2015} for a recent review). Although point source
searches of a number of blazars by IceCube have yield
non-detection\cite{Aartsen2014}, it is completely possible that a
fraction of IceCube neutrinos come from giant flare periods of
blazars, as may be the case for PKS B1424-418 and other
blazars\cite{Padovani2014}.  The large positional uncertainty of
this IC35 PeV neutrino is mainly due to its cascade nature. If a
muon-track PeV neutrino, with a typical position error of only one
degree, is identified to be in temporal and positional coincidence
with any blazar flares in future, the confidence of the
association will be significantly increased.

The accuracy of testing EP could be further improved if neutrinos
from GRBs could be detected in future, since the duration of GRBs
is much shorter than that of blazar flares.  The cosmic-ray proton
interactions with the GRB fireball photons are supposed to produce
a burst of neutrinos with energies up to  $\sim$ PeV (e.g.
ref.\cite{Waxman1997}; see refs.\cite{Meszaros2015} for a recent
review). With a typical duration of tens of seconds, the accuracy
could be improved by several orders of magnitude if a TeV-PeV
neutrino associated with the prompt burst is detected. Taking the
maximum time delay difference between the neutrino and the prompt
gamma-ray burst emission as $\delta t_\nu-\delta t_\gamma\leq {\rm
100 s}$, one would obtain $\gamma_\nu-\gamma_\gamma \leq 10^{-9}$
when the Shapiro delay due to the Virgo cluster is taken into
account. Similarly, the accuracy of constraining the LIV can be
significantly improved by GRB neutrinos.

We thank Xue-Feng Wu and He Gao for useful discussions. This work
is supported by the National Basic Research Program (¡°973¡±
Program) of China under grant 2014CB845800, the National Natural
Science Foundation  of  China under grant No. 11273016, and the
Basic Research Program of Jiangsu Province under Grant No.
BK2012011.


\begin{thebibliography}{srt}



\bibitem[Aartsen et al.(2013)]{IceCube2013} Aartsen, M.~G., Abbasi,
R., Ackermann, M., et al.\ 2013, \prd, 88, 112008


\bibitem[Aartsen et al.(2014a)]{IceCube2014} Aartsen, M.~G.,
Ackermann, M., Adams, J., et al.\ 2014a, Physical Review Letters,
113, 101101

\bibitem[Kadler et al.(2016)]{Kadler2016} Kadler, M., Krau{\ss},
F., Mannheim, K., et al.\ 2016,  Nature Physics, in press,
arXiv:1602.02012

\bibitem[Will (2006)]{Will} C. M. Will, Living Rev. Relativ. 9, 3 (2006); C. M. Will,
Living Rev. Relativ. 17, 4 (2014).

\bibitem[White et al.(1988)]{White1988} White, G.~L., Jauncey,
D.~L., Wright, A.~E., et al.\ 1988, \apj, 327, 561
\bibitem[Longo(1988)]{Longo1988} Longo, M.~J.\ 1988, Physical
Review Letters, 60, 173

\bibitem[Krauss
\& Tremaine(1988)]{Krauss&Tremaine1988} Krauss, L.~M., \&
Tremaine, S.\ 1988, Physical Review Letters, 60, 176

\bibitem[Gao et al.(2015)]{Gao2015} Gao, H., Wu, X.-F.,
\& M{\'e}sz{\'a}ros, P.\ 2015, \apj, 810, 121

\bibitem[Wei et al.(2015)]{Wei2015} Wei, J.-J., Gao, H., Wu,
X.-F., \& M{\'e}sz{\'a}ros, P.\ 2015, Physical Review Letters,
115, 261101

\bibitem[Wei et al.(2016)]{Wei2016} Wei J.-J., et al., 2016, ApJ, 818, L2.

\bibitem[Zhang (2016)]{Zhang2016}Zhang, S. N. 2016, arXiv:1601.04558

\bibitem[Nusser(2016)]{Nusser2016} Nusser, A.\ 2016,
arXiv:1601.03636
\bibitem[Wu et al. (2016)]{Wu2016} Wu, X. F. et al., 2016, arXiv:1602.01566
\bibitem[Kahya \& Desai (2016)]{Kahya2016} E. O. Kahya, S. Desai, 2016,  arXiv:1602.04779

\bibitem[Amelino-Camelia et al.(1998)]{Amelino-Camelia1998}
Amelino-Camelia, G., Ellis, J., Mavromatos, N.~E., Nanopoulos,
D.~V., \& Sarkar, S.\ 1998, \nat, 393, 763

\bibitem[Abdo et al.(2009)]{Abdo2009} Abdo, A.~A.,
Ackermann, M., Ajello, M., et al.\ 2009, \nat, 462, 331

\bibitem[Vasileiou et al.(2013)]{Vasileiou2013} Vasileiou, V.,
Jacholkowska, A., Piron, F., et al.\ 2013, \prd, 87, 122001

\bibitem[Ellis et al.(2008)]{Ellis2008} Ellis, J., Harries, N.,
Meregaglia, A., Rubbia, A., \& Sakharov, A.~S.\ 2008, \prd, 78,
033013

\bibitem[Jacob
\& Piran(2007)]{Jacob2007} Jacob, U., \& Piran, T.\ 2007, Nature
Physics, 3, 87



\bibitem[Aartsen et al.(2015)]{Aartsen2015} Aartsen, M.~G.,
Ackermann, M., Adams, J., et al. 2015, \apj, 805, L5

\bibitem[Shapiro(1964)]{Shapiro1964} Shapiro, I.~I.\ 1964, Physical
Review Letters, 13, 789

\bibitem[Mei et
al.(2007)]{Mei2007} Mei, S., Blakeslee, J.~P., C{\^o}t{\'e}, P.,
et al.\ 2007, \apj, 655, 144

\bibitem[Minakata
\& Smirnov(1996)]{Minakata1996PhRvD} Minakata, H., \& Smirnov,
A.~Y.\ 1996, \prd, 54, 3698

\bibitem[Fouqu{\'e} et
al.(2001)]{Fouque2001} Fouqu{\'e}, P., Solanes, J.~M., Sanchis,
T., \& Balkowski, C.\ 2001, A\&A, 375, 770
\bibitem[Lynden-Bell et al.(1988)]{Lynden-Bell1988} Lynden-Bell, D.,
Faber, S.~M., Burstein, D., et al.\ 1988, \apj, 326, 19


\bibitem[Cohen
\& Glashow(2011)]{Cohen2011} Cohen, A.~G., \& Glashow, S.~L.\
2011, Physical Review Letters, 107, 181803


\bibitem[Borriello et al. (2013)]{Borriello2013}Borriello, E.,  Chakraborty, S.,
Mirizzi A., Pasquale Dario Serpico, 2013, Physical Review D, vol.
87,  id. 116009
\bibitem[Stecker (2014)]{Stecker2014} F. W. Stecker, 2014, Astroparticle Physics, Volume 56, p. 16-18
\bibitem[Diaz et al. (2014)]{Diaz2014} J. S. Diaz, A. Kostelecky, M.
Mewes, 2014, \prd,  89,  043005

\bibitem[Planck Collaboration (2015)]{Planck2015} Planck Collaboration; Ade, P. A. R.; Aghanim, N.; Arnaud,
M. et al, 2015, eprint arXiv:1502.01589

\bibitem[Abramowski et al.(2011)]{Abramowski2011}
Abramowski, A., Acero, F., et al.\ 2011, Astroparticle Physics,
34, 738
\bibitem[Stecker et al. (1991)]{Stecker1991} Stecker, F. W., Done, C., Salamon, M. H., \& Sommers, P. 1991, \prl, 66,
2697

\bibitem [Mannheim et al. (1992)]{Mannheim1992} Mannheim, K., Stanev, T., \& Biermann, P. L., A\&A,
260, L1 (1992)
\bibitem[Murase(2015)]{Murase2015} Murase, K., 2015,
arXiv:1511.01590

\bibitem[Aartsen et al.(2014b)]{Aartsen2014} Aartsen, M.~G.,
Ackermann, M., Adams, J., et al.\ 2014b, \apj, 796, 109

\bibitem [Padovani \& Resconi (2014)]{Padovani2014}
 Padovani, P. \& Resconi, E., 2014, MNRAS, 443, 474-484








\bibitem[Waxman \& Bahcall (1997)]{Waxman1997}Waxman, E., \& Bahcall, J. 1997, \prl, 78, 2292

\bibitem[M{\'e}sz{\'a}ros (2015)]{Meszaros2015} M{\'e}sz{\'a}ros,P., 2015,  arXiv:1511.01396









\end{thebibliography}
\end{document}